\documentclass[draft,grl]{agutex}

\authorrunninghead{LIU ET AL.}

\titlerunninghead{Stability analysis of the Biot/squirt models}

\authoraddr{Corresponding author: W.-A. Yong,
Zhou Pei-Yuan Center for Appl. Math., Tsinghua Univ., Beijing 100084, China.
(wayong@tsinghua.edu.cn)}

\usepackage{amsmath}
\usepackage{CJK}
\usepackage{chemarrow}
\usepackage{setspace}
\usepackage{array}
\usepackage{graphicx}
\usepackage{subfigure}
\usepackage{multirow}
\usepackage{bm}
\usepackage{threeparttable}
\usepackage{lineno}

\newtheorem{Theorem}{Theorem}[section]
\newtheorem{proposition}{Proposition}[section]

\begin{document}

%
%

\title{Stability analysis of the Biot/squirt models for wave propagation in saturated porous media}
%
%

%
%



\authors{J.-W. Liu,\altaffilmark{1}
W.-A. Yong\altaffilmark{1}}


\altaffiltext{1}{Zhou Pei-Yuan Center for Appl. Math., Tsinghua Univ., Beijing 100084, China.}

%
%


\begin{abstract}
This work is concerned with the Biot/squirt (BISQ) models for wave propagation in saturated porous media. We show that the models allow exponentially exploding solutions, as time goes to infinity, when the characteristic squirt-flow coefficient is negative or has a non-zero imaginary part. We also show that the squirt-flow coefficient does have non-zero imaginary parts for some experimental parameters. Because the models are linear, the existence of such exploding solutions indicates instability of the BISQ models. This result calls on a reconsideration of the widely used BISQ theory.
Furthermore, we demonstrate that the 3D isotropic BISQ model is stable when the squirt-flow coefficient is positive. In particular, the original Biot model is unconditionally stable where the squirt-flow coefficient is 1.
\end{abstract}

%
%

%

\begin{article}

%
%

\section{Introduction}
The propagation of seismic waves in rocks with fluids has been among the most active research fields
in geoscience. Various theories [Biot, 1956a, b; Dvorkin et al., 1993; Mavko et al., 2009; M\"uller et al, 2010] focus on relating attenuation and dispersion of seismic waves in earth materials to physical properties of the rocks and fluids.
Attenuation means the exponential decay of wave amplitude with distance and dispersion is a variation of propagation velocity with frequency.
Among the various mechanisms related to attenuation and dispersion, the Biot and squirt-flow mechanisms are believed to be the most important ones [Dvorkin et al., 1993; Yang et al., 2002]. They have served as rigorous and formal foundations to study acoustic wave propagation in saturated porous media.

In Biot's theory [Biot, 1956a, b], pore fluids are forced to participate in the solid's motion due to inertia and viscosity.
The Biot mechanism is described through macroscopic rock properties and based on the macroscopic flow.
The squirt-flow mechanism is associated with the squirting of the pore fluid out of cracks as they are deformed by the compressional waves. It is related to microscopic rock properties and based on the local fluid flow [Dvorkin et al, 1993]. Clearly, the two mechanisms occur simultaneously and affect the process of seismic energy propagation and attenuation.

Dvorkin and Nur [1993] developed a consistent one-dimensional (1D) model, incorporating both
the Biot and squirt-flow mechanisms. That is the widely used Biot/Squirt-flow (BISQ) model. The BISQ theory directly relates the velocity and attenuation of P-waves to the macroscopic and microscopic parameters.
Later on, Parra [1997, 2000] extended the 1D model to the 3D case for a transversally isotropic poroelastic medium, which is based on the constitutive relations in the frequency domain given by Kazi-Aoual et al. [1988] and the isotropic description of the solid/fluid coupling effects developed by Biot [1956b]. Furthermore, Yang and Zhang [2002] adapted the BISQ theory for anisotropic porous media.
Up to now, it is well accepted [Mavko et al, 2009; M\"uller et al, 2010] that the BISQ theory is adequate to model acoustic wave propagation in saturated porous media. The acceptance is based mainly on experimental verifications of the predicted phase velocities or analysis of wave field snapshots for a short time. However, a mathematical study of the BISQ model does not seem available in the literature.

Observe that the models can be transformed to a system of first-order partial differential equations (see Section 3) and thereby possess the form of hyperbolic relaxation systems. Based on this, we are guided by the theory of hyperbolic relaxation systems developed in [Yong, 1992, 2001] and conduct a stability analysis of the BISQ model and its 3D isotropic extension. Our results show that the models allow spatially bounded but time exponentially exploding solutions when the characteristic squirt-flow coefficient is negative or has a non-zero imaginary part. By numerically examining some experimental parameters from [Dvorkin et al, 1993; King et al., 2000; Parra, 2000; Yang et al., 2011], we see that it is usual and realistic for the squirt-flow coefficient to have non-zero imaginary parts and for the models to have time exponentially exploding solutions. Because the models are linear, the existence of such exploding solutions indicates instability of the widely used models. This seems the first time to point out instability of the BISQ models.
Furthermore, we show that, when the squirt-flow coefficient is positive, the 3D isotropic model satisfies a structural stability condition for hyperbolic relaxation systems and therefore is stable. In particular, the original Biot model is unconditionally stable where the squirt-flow coefficient is 1.

Our results call on a reconsideration of the BISQ theory.
A possible modification would be to replace the squirt-flow coefficient with its real part.
The results may also provide reasonable explanations for some problematic numerical simulations reported in the literature. In [Parra, 2000] for a transversely isotropic medium, it was pointed out that the velocity dispersion and attenuation predicted by the BISQ model are not consistent with experimental data at the low angular frequency.
For the snapshot of seismic wave fields, Yang et al. [2002] found that it is hard to gain the slow P-waves by the BISQ model with the small frequency and the slow P-waves occur only after increasing the  frequency. In addition, the numerical results in Yang et al. [2002] indicate that the slow P-waves only occur when the fluid viscosity is negligible. On the other hand, we notice that the squirt-flow coefficient has always a nonzero imaginary part at small angular frequencies and is real when the viscosity vanishes. These coincide very well with our theoretical results: the BISQ model is unstable when the squirt-flow coefficient has a nonzero imaginary part and is stable when the squirt-flow coefficient is positive.

The paper is organized as follows. In Section 2 we show the existence of time-exponentially exploding solutions to the 1D BISQ model when the squirt-flow coefficient is negative or has a non-zero imaginary part. Here we also numerically examine some experimental parameters. Section 3 is devoted to the 3D isotropic extension, which is transformed to a system of first-order partial differential equations with the coefficient matrices given in Appendix. It is demonstrated that the instability result is valid for the 3D model and the positive squirt-flow coefficient ensures the structural stability of the extension. The final  section summarizes our main results.

\section{1D Model}
We start with the Biot/squirt (BISQ) dynamic equations for a two-component solid-fluid continuum in one dimension [Dvorkin et al., 1993]:
\begin{linenomath*}
\begin{eqnarray}\label{21}
&&(1-\phi)\rho_s u_{tt}+\phi\rho_f w_{tt}=Mu_{xx}-\alpha P_x,\nonumber\\[0.2cm]
&&\phi\rho_f w_{tt}-\rho_a(u_{tt}-w_{tt})-[\eta\phi^2(u_t-w_t)]/\kappa=-\phi P_x,\\[0.2cm]
&&P_t=-FS\left\{w_{xt}+[(\alpha-\phi)u_{xt}]/\phi\right\}.\nonumber
\end{eqnarray}
\end{linenomath*}
Here
$u=u(x, t)$ and $w=w(x, t)$ are the respective solid's and fluid's displacements at the position-time point $(x, t)$,
$P=P(x, t)$ is the total fluid pressure, $\phi\in(0,1)$ is the porosity of the solid, $M$ is the uniaxial modulus of the skeleton in drained conditions, $\alpha$ is the poroelastic coefficient, $\rho_s$, $\rho_f$ and $\rho_a$ are the respective solid's, fluid's and  additional coupling densities, $\eta$ is the viscosity of the fluid, $\kappa$ is the permeability of the skeleton, and $F$ is the Biot-flow coefficient. All these parameters are positive (see Table 1 below). Subscripts $x$ and $t$ indicate partial derivatives ($P_x=\frac{\partial P}{\partial x}$, $u_{tt}=\frac{\partial^2 u}{\partial t^2}$, etc.).

In (1), $S$ is the characteristic squirt-flow coefficient. It has the expression
\begin{linenomath*}
\begin{eqnarray}\label{22}
S=1-[2J_1(\Lambda R)]/[\Lambda R J_0(\Lambda R)]
\end{eqnarray}
\end{linenomath*}
with
\begin{linenomath*}
\begin{eqnarray}\label{23}
\Lambda^2=\left\{\rho_f f^2\left[1+ \rho_a/(\phi\rho_f)+\sqrt{-1}(\eta \phi)/(\kappa \rho_f f)\right]\right\}/F,
\end{eqnarray}
\end{linenomath*}
where $J_0$ and $J_1$ the respective Bessel functions of zero- and first-order, R is the characteristic squirt-flow length, and $f$ is the angular frequency.  Observe that $\Lambda$ and thereby $S$ are complex in general. See also Table 1 below.

By substituting the third  equation in (1) into the first two equations therein, we get two equations for the displacements $u$ and $w$. For the resultant two equations, we look for their solutions of the form [Yong, 1992]
\begin{linenomath*}
\begin{equation*}
u(x,t)=u(t)\mathrm{exp}(i\xi x),~~~~~~w(x,t)=w(t)\mathrm{exp}(i\xi x)
\end{equation*}
\end{linenomath*}
with $\xi$ real. In this way, we obtain
\begin{linenomath*}
\begin{eqnarray}\label{24}
&&\rho_1u_{tt}+\rho_2w_{tt}=-\left\{M+[\alpha FS(\alpha-\phi)]/\phi\right\}\xi^2u-\alpha FS\xi^2w,\nonumber\\
[0.4cm]
&&\rho_{12}u_{tt}+\rho_{22}w_{tt}-[\eta\phi^2(u_t-w_t)]/\kappa=\phi FS\xi^2\left\{-[(\alpha-\phi)u]/\phi-w\right\}.
\end{eqnarray}
\end{linenomath*}
Here $\rho_1=(1-\phi)\rho_s$, $\rho_2=\phi\rho_f$, $\rho_{12}=-\rho_a$ and $\rho_{22}=\phi\rho_f+\rho_a$.

Set $v=u_t$ and $V=w_t$. The system of ordinary differential equations \eqref{24} can be rewritten as
\begin{linenomath*}
\begin{eqnarray*}
&&u_t=v,\\
[0.4cm]
&&w_t=V,\\
[0.4cm]
&&\rho_1v_t+\rho_2V_t=-\left\{M+[\alpha FS(\alpha-\phi)]/\phi\right\}\xi^2u-\alpha FS\xi^2w,\\
[0.4cm]
&&\rho_{12}v_t+\rho_{22}V_t=-\xi^2FS(\alpha-\phi)u-\xi^2\phi FSw+(\eta\phi^2v)/\kappa-(\eta\phi^2V)/\kappa.
\end{eqnarray*}
\end{linenomath*}
This can be further rewritten as
\begin{linenomath*}
\begin{equation}\label{25}
\bm{A}\bm{U}_t=\bm{B}\bm{U}.
\end{equation}
\end{linenomath*}
Here $\bm{U}=[u, w, v, V]^T$ is an unknown vector-valued function of $t\in [0,+\infty)$, $\bm{A}$ and $\bm{B}$ are constant coefficient matrices
\begin{linenomath*}
\begin{eqnarray*}
~~~\bm{A}=\left[
\begin{array}{cccc}
\vspace{0.2cm}
1&0&0&0\\
\vspace{0.2cm}
0&1&0&0\\
\vspace{0.2cm}
0&0&\rho_1&\rho_2\\
0&0&\rho_{12}&\rho_{22}\\
\end{array}
\right],
\quad
\bm{B}=\left[
\begin{array}{cccc}
\vspace{0.2cm}
0&0&1&0\\
\vspace{0.2cm}
0&0&0&1\\
\vspace{0.2cm}
-\xi^2\left[M+\alpha FS(\alpha-\phi)/\phi\right)]&-\xi^2\alpha FS&0&0\\
-\xi^2 FS(\alpha-\phi)&-\xi^2\phi FS&(\eta\phi^2)/\kappa&-(\eta\phi^2)/\kappa\\
\end{array}
\right]
.
\end{eqnarray*}
\end{linenomath*}
The solution to the system \eqref{25} with initial data $\bm{U}_0$ is
\begin{linenomath*}
\begin{equation*}
\bm{U}(t)=\rm{exp}(t\bf{A}^{-1}\bm{B})\bm{U}_0.
\end{equation*}
\end{linenomath*}
This solution will exponentially explode as $t$ goes to infinity, provided that the matrix $\bm{A}^{-1}\bm{B}$ has an eigenvalue with positive real part and $\bf{U}$$_0$ is the corresponding eigenvector. Because the BISQ model (1) is linear, the existence of such exploding solutions means instability.

The eigenvalues of matrix $\bm{A}^{-1}\bm{B}$ are zeros of its characteristic polynomial
\begin{linenomath*}
\begin{equation}\label{26}
\lambda^4+(\eta b\lambda^3)/\kappa+\xi^2c\lambda^2+(\eta\xi^2d\lambda)/\kappa+\xi^4e=0
\end{equation}
\end{linenomath*}
with coefficients
\begin{linenomath*}
\begin{eqnarray}\label{27}
&&b=[\phi^2(\rho_1+\rho_2)]/(\rho_1\rho_{22}-\rho_2\rho_{12}),\nonumber\\[4mm]
&&c=\left\{M\phi\rho_{22}+[(\rho_1+\rho_2)\phi^2-2\rho_2\alpha\phi+\rho_{22}\alpha^2] FS\right\}/\left[\phi(\rho_1\rho_{22}-\rho_2\rho_{12})\right],\nonumber\\[4mm]
&&d=\left[\phi\left(M\phi+\alpha^2 FS\right)\right]/(\rho_1\rho_{22}-\rho_2\rho_{12}),\qquad
e=(M\phi FS)/(\rho_1\rho_{22}-\rho_2\rho_{12}).
\end{eqnarray}
\end{linenomath*}
These coefficients are derived through a direct calculation and also with Mathlab. Notice that $(\rho_1\rho_{22}-\rho_2\rho_{12})$ is always positive for $\rho_{12}<0$.

About the matrix $\bm{A}^{-1}\bm{B}$, we have

\begin{proposition}
If $S<0$, then the matrix $\bm{A}^{-1}\bm{B}$ with $\xi\ne0$ has eigenvalues with positive real parts.
\end{proposition}

$Proof$. In case $S$ is real, the coefficients of the characteristic polynomial are all real. Thus, the complex eigenvalues must be conjugate. If $S<0$, the product of four eigenvalues $\xi^4e$ is negative. This means that there exists at least  two real eigenvalues with opposite sign. Hence the matrix $\bm{A}^{-1}\bm{B}$ has eigenvalues with positive real parts.

Recall that the squirt-flow coefficient $S$ is complex in general. We point out the following important fact. In what follows, we denote by Re($a$) and Im($a$) the real and imaginary parts of complex number $a$, respectively.

\begin{Theorem}
If the imaginary part of $S$ is non-zero, then the matrix $\bm{A}^{-1}\bm{B}$ has eigenvalues with positive real parts whenever $\xi\kappa/\eta$ is sufficiently large (positive or negative).
\end{Theorem}
$Proof$.
Observe that the roots of the algebraic equation \eqref{26} can be written as
\begin{linenomath*}
$$
\lambda=\xi\tilde\lambda\left[\eta/(\xi\kappa)\right]
$$
\end{linenomath*}
with $\tilde\lambda=\tilde\lambda(\epsilon)$ solving
\begin{linenomath*}
$$
{\tilde\lambda}^4 + \epsilon b{\tilde\lambda}^3 + c{\tilde\lambda}^2 + \epsilon d\tilde\lambda + e=0.
$$
\end{linenomath*}
Note that $\tilde\lambda(\epsilon)$ is continuous with respect to $\epsilon$ [Kato, 1982].  Then
$\mu=\lim_{\epsilon\rightarrow0}\tilde\lambda(\epsilon)$ solves
\begin{linenomath*}
$$
\mu^4+c\mu^2+e=0.
$$
\end{linenomath*}
Denote by $\mu_j (j=1,2,3,4)$ the roots of this algebraic equation. Because their product $e$ is complex whenever Im(S)$\neq 0$, there must be a $j_1\in\{1,2,3,4\}$ such that Re$(\mu_{j_1})\ne 0$.
Since $\sum_j\mu_j=0$, there must be another $j_2\in\{1,2,3,4\}$ such that Re$(\mu_{j_1})$Re$(\mu_{j_2})< 0$. Thus, by the continuity argument the algebraic equation \eqref{26} has eigenvalues with positive real parts for large $\xi\kappa/\eta$. This completes the proof.

This theorem can be strengthened as follows.

\begin{Theorem}
Under the condition
\begin{linenomath*}
$$
|Im(c)||\xi|\kappa>\sqrt{2}b\eta\sqrt{\max\{0, Re(c)\}},
$$
\end{linenomath*}
the matrix $\bm{A}^{-1}\bm{B}$ has eigenvalues with positive real parts.
\end{Theorem}
$Proof$.
Let $\lambda_j=x_j+\sqrt{-1}y_j (j=1,2,3,4)$ be solutions to the algebraic equation \eqref{26}. Notice that $b=[\phi^2(\rho_1+\rho_2)]/(\rho_1\rho_{22}-\rho_2\rho_{12})>0$. We have $\sum\limits_{j}x_j=-b\eta/\kappa, \sum\limits_{j}y_j=0$ and thereby $-2\sum_{j< k}y_j y_k=y_1^2+y_2^2 + y_3^2+y_4^2$.

If $x_j\le 0$ for all $j$, we deduce that
\begin{linenomath*}
\begin{eqnarray*}
2\xi^2Re(c)=2Re\sum_{j< k}\lambda_j \lambda_k =2\sum_{j< k}(x_j x_k-y_j y_k)=2\sum_{j< k}x_j x_k +\sum_j y_j^2\ge \sum_j y_j^2
\end{eqnarray*}
\end{linenomath*}
and
\begin{linenomath*}
\begin{eqnarray*}
&&\xi^2|Im(c)|=|Im \sum_{j< k}\lambda_j \lambda_k| = |\sum_{j< k}(x_j y_k+x_k y_j)| = |\sum_{j\neq k} x_j y_k| = |\sum_{j}x_j y_j|\\[4mm]
\le && \sqrt{\sum_{j}x_j^2}\sqrt{\sum_{j}y_j^2}\le |\sum_{j}x_j|\sqrt{2}|\xi|\sqrt{Re(c)}
= \sqrt{2}(b\eta/\kappa)|\xi|\sqrt{Re(c)}.
\end{eqnarray*}
\end{linenomath*}
Hence, the matrix $\bm{A}^{-1}\bm{B}$ has eigenvalues with positive real parts under the condition of the theorem. This completes the proof.

The above results indicate that the 1D BISQ model (1) allows spatially bounded but time-exponentially exploding solutions when the squirt-flow coefficient is negative or has a non-zero imaginary part. Theorem 2.2 gives a sufficient condition for the existence of the exploding solutions.

Motivated by the above results, we compute the eigenvalues with parameters from Refs. [Dvorkin et al, 1993; King et al, 2000; Parra, 2000; Yang et al., 2011] and $\xi=1$. The following table contains the imaginary parts of $FS$ and the corresponding eigenvalues with the largest real parts, which are all positive.
%
\begin{linenomath*}
$$\rm{\bm{Table\;1.}\quad Eigenvalues\; with\;experimental\;parameters\;and}\;\xi=1$$
\end{linenomath*}

\begin{tabular}{l c c c c}
\hline
{Data Sources}&[Dvorkin et al., 1993]&[King et al., 2000]&[Parra, 2000]&[Yang et al., 2011]\\
\hline
$M~~(GPa)$&35.480&20.217 &23.278&29\\
$\phi$~~($\%$)& 15&27.4 &23&15\\
  $\alpha$& 0.5789&0.7692 &0.6197&0.5702\\
$\rho_s$  ($kg/m^3$)&2650&2550&2750& 2600\\
 $\rho_f$  ($kg/m^3$)&1000&1000&1000&1000\\
  $\rho_a$  ($kg/m^3$)&420&420&420& 450\\
 $\eta/\kappa$ ($kg/m^4\cdot s$)& $4\times 10^4$&5$\times 10^4$&8$\times 10^4$&100\\
 $f$  (KHz)& 50&600&50& 50\\
 $R$  (m)& 0.001&0.015&0.001& 0.001\\
{Im(FS)}&-37.5587&-2.049$\times 10^4$&-115.1&-0.0939\\
Max[Re($\lambda$)]&16.9131&6.018$\times 10^{-13}$&14.7252&17.6842\\
 \hline
\end{tabular}


\section{3D isotropic models}

The poroelastic wave equations for a 3-D motion including the Biot and squirt-flow mechanisms in an isotropic medium read as [Parra, 1997; Yang et al, 2002]
\begin{linenomath*}
\begin{eqnarray}\label{31}
&&\mu\nabla\nabla\cdot \bm{u} +\mu\Delta \bm{u} +\left\{\lambda+[\alpha F S(\alpha-\phi)]/\phi\right\}\nabla\nabla\cdot \bm{u} + \alpha F S\nabla\nabla\cdot \bm{w}
=\rho_1\frac{\partial^2\bm{u}}{\partial t^2} +\rho_2\frac{\partial^2\bm{w}}{\partial t^2},\nonumber\\[0.5cm]
&&FS(\alpha-\phi)\nabla\nabla\cdot \bm{u} + \phi FS\nabla\nabla\cdot \bm{w}
=\rho_{12}\frac{\partial^2\bm{u}}{\partial t^2} +\rho_{22}\frac{\partial^2\bm{w}}{\partial t^2}
+[\eta\phi^2\left(\frac{\partial \bm{w}}{\partial t}-\frac{\partial \bm{u}}{\partial t}\right)]/\kappa.
\vspace{0.3cm}
\end{eqnarray}
\end{linenomath*}
Here $\lambda$ and $\mu$ are the Lame coefficients, $\bm{u}=\bm{u}(x_1, x_2, x_3, t)$ and $\bm{w}=\bm{w}(x_1, x_2, x_3, t)$ denote the respective solid's and fluid's displacement vectors, and the other parameters are same as those in (1) with $M=\lambda+2\mu$.

Let $[u(x, t), w(x, t)]$ solve the equations in (1). It is direct to verify that
\begin{linenomath*}
$$
\bm{u}(x_1, x_2, x_3, t) =\left[
\begin{array}{c}
1\\
0\\
0
\end{array}
\right]u(x_1, t) \quad \bm{w}(x_1, x_2, x_3, t)=\left[
\begin{array}{c}
1\\
0\\
0
\end{array}
\right]w(x_1, t)
$$
\end{linenomath*}
is a solution to equations \eqref{31}. Thus, we have

\begin{proposition}
If $S<0$ or Im$(S)\ne0$, then equations \eqref{31} have spatially bounded but time-exponentially exploding solutions.
\end{proposition}

Next we study the case $S>0$, including the Biot model where $S=1$. First of all, we transform the second-order equations \eqref{31} into a system of first-order partial differential equations. To do this, we write
\begin{linenomath*}
$$
\bm{u}=\left[
\begin{array}{c}
u_1\\
u_2\\
u_3
\end{array}
\right], \qquad \bm{w}=\left[
\begin{array}{c}
w_1\\
w_2\\
w_3
\end{array}
\right]
$$
\end{linenomath*}
and set
\begin{linenomath*}
\begin{equation}\label{32}
\begin{array}{l}
W_i=\frac{\partial u_i}{\partial x_i}~,~
v_i=\frac{\partial u_i}{\partial t}~,~V_i=\frac{\partial w_i}{\partial t}~,~T_{ij}=\frac{\partial u_i}{\partial x_j}+\frac{\partial u_j}{\partial x_i}~(i< j)~ ,~L=FS\left(\sum\limits_{j=1}^3\frac{\partial w_j}{\partial x_j}\right)
\end{array}
\end{equation}
\end{linenomath*}
for $i, j =1, 2, 3$.
With such notations, the linear system \eqref{31} can be rewritten as
\begin{linenomath*}
\begin{eqnarray}\label{33}
&\rho_1\frac{\partial}{\partial t}v_1+\rho_2\frac{\partial}{\partial t}V_1- \left\{\lambda+[\alpha FS(\alpha-\phi)]/\phi\right\}\frac{\partial}{\partial x_1}(\sum\limits_{j=1}^3 W_j)
-2\mu \frac{\partial}{\partial x_1}W_1-\mu\frac{\partial}{\partial x_2}T_{12}-\mu\frac{\partial}{\partial x_3}T_{13}-\alpha\frac{\partial}{\partial x_1}L=0,\nonumber\\
[0.4cm]
&\rho_1\frac{\partial}{\partial t}v_2+\rho_2\frac{\partial}{\partial t}V_2-\left\{\lambda+[\alpha FS(\alpha-\phi)]/\phi\right\}\frac{\partial}{\partial x_2}(\sum\limits_{j=1}^3 W_j)-2\mu \frac{\partial}{\partial x_2}W_2-\mu\frac{\partial}{\partial x_1}T_{12}-\mu\frac{\partial}{\partial x_3}T_{23}-\alpha\frac{\partial}{\partial x_2}L=0,\nonumber\\
[0.4cm]
&\rho_1\frac{\partial}{\partial t}v_3+\rho_2\frac{\partial}{\partial t}V_3-\left\{\lambda+[\alpha FS(\alpha-\phi)]/\phi\right\}\frac{\partial}{\partial x_3}(\sum\limits_{j=1}^3 W_j)-2\mu\frac{\partial}{\partial x_3}W_3
-\mu\frac{\partial}{\partial x_1}T_{13}-\mu\frac{\partial}{\partial x_2}T_{23}-\alpha\frac{\partial}{\partial x_3}L=0,\nonumber\\[0.4cm]
&\rho_{12}\frac{\partial}{\partial t}\bm{v}+\rho_{22}\frac{\partial}{\partial t}\bm{V}-FS(\alpha-\phi)\nabla(\sum\limits_{j=1}^3 W_j)-\phi\nabla L
=-[\eta\phi^2\left(\bm{V}-\bm{v}\right)]/\kappa.
\end{eqnarray}
\end{linenomath*}
In addition, by definition it is easy to see that
\begin{linenomath*}
\begin{equation}\label{34}
\begin{array}{l}
\frac{\partial W_i}{\partial t}-\frac{\partial v_i}{\partial x_i}=0~,~
\frac{\partial T_{ij}}{\partial t}-\frac{\partial v_i}{\partial x_j}-\frac{\partial v_j}{\partial x_i}=0~(i< j)~ ,~\frac{\partial L}{\partial t}-FS\left(\sum\limits_{j=1}^3\frac{\partial V_j}{\partial x_j}\right)=0.
\end{array}
\end{equation}
\end{linenomath*}
In \eqref{33} and \eqref{34} there are 13 first-order partial differential equations. Clearly, they can be written in the vector form:
\begin{linenomath*}
\begin{equation}\label{35}
\bm{A}\frac{\partial}{\partial t}\bm{U}+\sum\limits_{j=1}^3\bm{A}_{j}\frac{\partial}{\partial x_j}\bm{U}=\bm{B} \bm{U}
\end{equation}
\end{linenomath*}
with $\bm{U}= [W_1,W_2,W_3,T_{12},T_{13},T_{23},L,v_{1},V_{1},v_{2},V_{2},v_3,V_3]^T$ and the coefficient matrices given in Appendix. Note that $\bm{A}$ is invertible.

This system of first-order partial differential equations possesses the form of hyperbolic relaxation systems, whose fundamental properties were systematically identified in [Yong, 2001]. Here we show that the system \eqref{35} with $S>0$ satisfies the second structural stability condition in [Yong, 1992, 2001]. More precisely, we have

\begin{Theorem}
If $S>0$, then there exists a symmetric positive definite matrix $\bm{A}_0$ such that the matrices
$\bm{A}^T\bm{A}_0\bm{A}_{j}$ ($j=1, 2, 3$) are symmetric and $\bm{A}^T\bm{A}_0\bm{B}$ is symmetric nonpositive definite. Moreover, the $L^2$-norm of the solutions to the system \eqref{35} are uniformly bounded in time.
\end{Theorem}
$Proof$. The symmetric positive definite matrix $\bm{A}_0$ is given in Appendix. One can directly verify that $\bm{A}^T\bm{A}_0\bm{A}_{j}$ ($j=1, 2, 3$) are symmetric and $\bm{A}^T\bm{A}_0\bm{B}$ is symmetric nonpositive definite.

To show the uniform boundedness, we multiply the system \eqref{35} with $\bm{U}^T\bm{A}^T\bm{A}_0$ and use the symmetry of $\bm{A}^T\bm{A}_0\bm{A}_{j}$ ($j=1, 2, 3$) to obtain
\begin{linenomath*}
$$
\frac{\partial}{\partial t}[\bm{U}^T\bm{A}^T\bm{A}_0\bm{A}\bm{U}] +
\sum_j\frac{\partial}{\partial x_j}[\bm{U}^T\bm{A}^T\bm{A}_0\bm{A}_j\bm{U}] =2\bm{U}^T\bm{A}^T\bm{A}_0\bm{B}\bm{U} .
$$
\end{linenomath*}
Integrating this equation over $[x_1, x_2, x_3]\in R^3$ gives
\begin{linenomath*}
$$
\frac{\partial}{\partial t}\int_{R^3} \bm{U}^T\bm{A}^T\bm{A}_0\bm{A}\bm{U} dx_1dx_2dx_3= 2\int_{R^3} \bm{U}^T\bm{A}^T\bm{A}_0\bm{B}\bm{U} dx_1dx_2dx_3\le0,
$$
\end{linenomath*}
where the inequality is due to the nonpositive definiteness of the symmetric matrix $\bm{A}^T\bm{A}_0\bm{B}$. Let the largest and smallest eigenvalues of the symmetric positive definite matrix $\bm{A}^T\bm{A}_0\bm{A}$ be $A$ and $a$. Then it follows from the last inequality that
\begin{linenomath*}
\begin{eqnarray*}
\vspace{0.4cm}
\qquad\qquad\qquad\qquad& \int_{R^3} \bm{U}^T(x_1, x_2, x_3, t)\bm{U}(x_1, x_2, x_3, t) dx_1dx_2dx_3\\
\vspace{0.4cm}
\le
& a^{-1}\int_{R^3} \bm{U}^T(x_1, x_2, x_3, t)\bm{A}^T\bm{A}_0\bm{A}\bm{U}(x_1, x_2, x_3, t) dx_1dx_2dx_3\\
\vspace{0.4cm}
\le & a^{-1}\int_{R^3} \bm{U}^T(x_1, x_2, x_3, 0)\bm{A}^T\bm{A}_0\bm{A}\bm{U}(x_1, x_2, x_3, 0) dx_1dx_2dx_3\\
\vspace{0.4cm}
\le & a^{-1}A\int_{R^3} \bm{U}^T(x_1, x_2, x_3, 0)\bm{U} (x_1, x_2, x_3, 0) dx_1dx_2dx_3,
\end{eqnarray*}
\end{linenomath*}
namely,
\begin{linenomath*}
$$
\|\bm{U}(\cdot, t)\|^2_{L^2}\le a^{-1}A\|\bm{U}(\cdot, 0)\|^2_{L^2}
$$
\end{linenomath*}
for all $t>0$. This completes the proof.

The last theorem indicates the stability of the 3D isotropic BISQ model \eqref{31} with any parameters satisfying $S>0$. Particularly, the original Biot model [Biot, 1956a, b] as the high-frequency limit of the 3D BISQ model \eqref{31} is unconditionally stable. On the other hand, we can directly verify the stability of the Biot model when $R(2N+A)-Q^2>0$ (Biot's parameters). The latter was identified also by Biot in [Biot, 1956a] with physical considerations, that is, it is equivalent to the positiveness  of the kinetic
energy.

\section{Conclusions}
In this work, we conduct a critical assessment of the BISQ models for isotropic poroelastic media and obtain the following conclusions.

\begin{itemize}

\item When the squirt-flow coefficient $S$ is positive, the 3D isotropic BISQ model is stable. Particularly, the original Biot model [Biot, 1956a, b] is unconditionally stable.

\item Otherwise, the model allows spatially bounded but time-exponentially exploding solutions and therefore are unstable.

\item Theorem 2.2 gives a sufficient condition for the existence of the exploding solutions, which requires  that $\xi\kappa/\eta$ is large enough when Im$(S)\ne 0$. Table 1 further shows that it is realistic and usual to have Im$(S)\ne 0$ and eigenvalues with positive real parts.

\end{itemize}
These, for the first time, reveal the instability of the BISQ models and call on a reconsideration of the widely used BISQ theory.


%
%
\section{Appendix. Coefficient matrices for the 3D isotropic model}

In Equation \eqref{35}, the coefficient matrices are
\begin{linenomath*}
\begin{equation*}
\bm{B}=-[(\eta\phi^2)/\kappa]
\left[
\begin{array}{ccccccccccccc}
\begin{smallmatrix}
0&0&0&0&0&0&0&0&0&0&0&0&0\\
0&0&0&0&0&0&0&0&0&0&0&0&0\\
0&0&0&0&0&0&0&0&0&0&0&0&0\\
0&0&0&0&0&0&0&0&0&0&0&0&0\\
0&0&0&0&0&0&0&0&0&0&0&0&0\\
0&0&0&0&0&0&0&0&0&0&0&0&0\\
0&0&0&0&0&0&0&0&0&0&0&0&0\\
0&0&0&0&0&0&0&0&0&0&0&0&0\\
0&0&0&0&0&0&0&-1&1&0&0&0&0\\
0&0&0&0&0&0&0&0&0&0&0&0&0\\
0&0&0&0&0&0&0&0&0&-1&1&0&0\\
0&0&0&0&0&0&0&0&0&0&0&0&0\\
0&0&0&0&0&0&0&0&0&0&0&-1&1
\end{smallmatrix}
\end{array}
\right],
\end{equation*}
\end{linenomath*}

\begin{linenomath*}
\begin{equation*}
\bm{A}=
\left[
\begin{array}{ccccccccccccc}
\begin{smallmatrix}
1&0&0&0&0&0&0&0&0&0&0&0&0\\
0&1&0&0&0&0&0&0&0&0&0&0&0\\
0&0&1&0&0&0&0&0&0&0&0&0&0\\
0&0&0&1&0&0&0&0&0&0&0&0&0\\
0&0&0&0&1&0&0&0&0&0&0&0&0\\
0&0&0&0&0&1&0&0&0&0&0&0&0\\
0&0&0&0&0&0&1&0&0&0&0&0&0\\
0&0&0&0&0&0&0&\rho_{1}&\rho_{2}&0&0&0&0\\
0&0&0&0&0&0&0&\rho_{12}&\rho_{22}&0&0&0&0\\
0&0&0&0&0&0&0&0&0&\rho_{1}&\rho_{2}&0&0\\
0&0&0&0&0&0&0&0&0&\rho_{12}&\rho_{22}&0&0\\
0&0&0&0&0&0&0&0&0&0&0&\rho_{1}&\rho_{2}\\
0&0&0&0&0&0&0&0&0&0&0&\rho_{12}&\rho_{22}
\end{smallmatrix}
\end{array}
\right],
\end{equation*}
\end{linenomath*}

\begin{linenomath*}
\begin{equation*}
\begin{split}
\bm{A}_1=
\left[
\begin{array}{ccccccccccccc}
\begin{smallmatrix}
0&0&0&0&0&0&0&-1&0&0&0&0&0\\
0&0&0&0&0&0&0&0&0&0&0&0&0\\
0&0&0&0&0&0&0&0&0&0&0&0&0\\
0&0&0&0&0&0&0&0&0&-1&0&0&0\\
0&0&0&0&0&0&0&0&0&0&0&-1&0\\
0&0&0&0&0&0&0&0&0&0&0&0&0\\
0&0&0&0&0&0&0&0&-FS&0&0&0&0\\
-A_1&
-A_2&
-A_2&
0&0&0&-\alpha&0&0&0&0&0&0\\
-A_3&
-A_3&
-A_3&0&0&0&-\phi&0&0&0&0&0&0\\
0&0&0&-\mu&0&0&0&0&0&0&0&0&0\\
0&0&0&0&0&0&0&0&0&0&0&0&0\\
0&0&0&0&-\mu&0&0&0&0&0&0&0&0\\
0&0&0&0&0&0&0&0&0&0&0&0&0
\end{smallmatrix}
\end{array}
\right],
\end{split}
\end{equation*}
\end{linenomath*}

\begin{linenomath*}
\begin{equation*}
\bm{A}_2=
\left[
\begin{array}{ccccccccccccc}
\begin{smallmatrix}
0&0&0&0&0&0&0&0&0&0&0&0&0\\
0&0&0&0&0&0&0&0&0&-1&0&0&0\\
0&0&0&0&0&0&0&0&0&0&0&0&0\\
0&0&0&0&0&0&0&-1&0&0&0&0&0\\
0&0&0&0&0&0&0&0&0&0&0&0&0\\
0&0&0&0&0&0&0&0&0&0&0&-1&0\\
0&0&0&0&0&0&0&0&0&0&-F S&0&0\\
0&0&0&-\mu&0&0&0&0&0&0&0&0&0\\
0&0&0&0&0&0&0&0&0&0&0&0&0\\
-A_2&
-A_1&
-A_2&
0&0&0&-\alpha&0&0&0&0&0&0\\
-A_3&
-A_3&
-A_3&0&0&0&-\phi&0&0&0&0&0&0\\
0&0&0&0&0&-\mu&0&0&0&0&0&0&0\\
0&0&0&0&0&0&0&0&0&0&0&0&0
\end{smallmatrix}
\end{array}
\right],
\end{equation*}
\end{linenomath*}

\begin{linenomath*}
\begin{equation*}
\bm{A}_3=
\left[
\begin{array}{ccccccccccccc}
\begin{smallmatrix}
0&0&0&0&0&0&0&0&0&0&0&0&0\\
0&0&0&0&0&0&0&0&0&0&0&0&0\\
0&0&0&0&0&0&0&0&0&0&0&-1&0\\
0&0&0&0&0&0&0&0&0&0&0&0&0\\
0&0&0&0&0&0&0&-1&0&0&0&0&0\\
0&0&0&0&0&0&0&0&0&-1&0&0&0\\
0&0&0&0&0&0&0&0&0&0&0&0&-F S\\
0&0&0&0&-\mu&0&0&0&0&0&0&0&0\\
0&0&0&0&0&0&0&0&0&0&0&0&0\\
0&0&0&0&0&-\mu&0&0&0&0&0&0&0\\
0&0&0&0&0&0&0&0&0&0&0&0&0\\
-A_2&
-A_2&
-A_1&
0&0&0&-\alpha&0&0&0&0&0&0\\
-A_3&
-A_3&
-A_3&0&0&0&-\phi&0&0&0&0&0&0
\end{smallmatrix}
\end{array}
\right],
\end{equation*}
\end{linenomath*}
where $A_1=2\mu+A_2,~A_2=\lambda+[\alpha FS(\alpha-\phi)]/\phi,~A_3=FS(\alpha-\phi).$

The matrix $\bm{A}_0$ in Theorem 3.1 is constructed as
\begin{linenomath*}
\begin{equation*}
\begin{split}
\bm{A}_0=
\left[
\begin{array}{ccccccccccccc}
\begin{smallmatrix}
A_4&A_5&A_5&0&0&0&A_6&0&0&0&0&0&0\\
A_5&A_4&A_5&0&0&0&A_6&0&0&0&0&0&0\\
A_5&A_5&A_4&0&0&0&A_6&0&0&0&0&0&0\\
0&0&0&\mu/(\rho_1+\rho_2)&0&0&0&0&0&0&0&0&0\\
0&0&0&0&\mu/(\rho_1+\rho_2)&0&0&0&0&0&0&0&0\\
0&0&0&0&0&\mu/(\rho_1+\rho_2)&0&0&0&0&0&0&0\\
A_6&A_6&A_6&0&0&0&
\phi/[FS(\rho_1+\rho_2)]&0&0&0&0&0&0\\
0&0&0&0&0&0&0&A_8&
-A_7&0&0&0&0\\
0&0&0&0&0&0&0&-A_7&(\rho_1\rho_{22}-\rho_2\rho_{12})^{-1}&0&0&0&0\\
0&0&0&0&0&0&0&0&0&A_8&-A_7&0&0\\
0&0&0&0&0&0&0&0&0&-A_7&(\rho_1\rho_{22}-\rho_2\rho_{12})^{-1}&0&0\\
0&0&0&0&0&0&0&0&0&0&0&A_8&-A_7\\
0&0&0&0&0&0&0&0&0&0&0&-A_7&(\rho_1\rho_{22}-\rho_2\rho_{12})^{-1}
\end{smallmatrix}
\end{array}
\right]
\end{split}
\end{equation*}
\end{linenomath*}
with
\begin{linenomath*}
\begin{eqnarray*}
&&A_4=\{\lambda+2\mu+[FS(\alpha-\phi)^2]/\phi\}/(\rho_1+\rho_2),~
\quad A_5=\{\lambda+[FS(\alpha-\phi)^2]/\phi\}/(\rho_1+\rho_2),~
\\[4mm]
&&A_6=(\alpha-\phi)/(\rho_1+\rho_2),~
\qquad\qquad\qquad\qquad\qquad A_7= (\rho_{12}+\rho_{22})/[(\rho_1+\rho_2)(\rho_1\rho_{22}-\rho_2\rho_{12})],
\\[4mm]
&&A_8=(\rho_1+\rho_2)^{-2}
+(\rho_{12}+\rho_{22})^2/[(\rho_1+\rho_2)^2(\rho_1\rho_{22}-\rho_2\rho_{12})].
\end{eqnarray*}
\end{linenomath*}
Recall that $S>0$ and $\rho_1\rho_{22}-\rho_2\rho_{12}>0$ for $\rho_{12}<0$. It can be directly checked that the matrix $\bm{A}_0$ is symmetric and positive definite.
\end{article}
%
%
%
%
%
%
%
%


\end{document}